\documentclass[twocolumn,prl,preprintnumbers,superscriptaddress,showpacs]{revtex4}

\usepackage{graphicx}
\usepackage{amsmath,amsfonts,amssymb,amsxtra}
\usepackage{units}
\usepackage{color}
\usepackage{textcomp}
\usepackage{amsmath}
\usepackage{exscale}

\makeatletter
\DeclareRobustCommand{\chemical}[1]{%
  {\(\m@th
   \edef\resetfontdimens{\noexpand\)%
       \fontdimen16\textfont2=\the\fontdimen16\textfont2
       \fontdimen17\textfont2=\the\fontdimen17\textfont2\relax}%
   \fontdimen16\textfont2=2.7pt \fontdimen17\textfont2=2.7pt
   \mathrm{#1}%
   \resetfontdimens}}
\DeclareRobustCommand{\bchemical}[1]{%
  {\(\m@th
   \edef\resetfontdimens{\noexpand\)%
       \fontdimen16\textfont2=\the\fontdimen16\textfont2
       \fontdimen17\textfont2=\the\fontdimen17\textfont2\relax}%
   \fontdimen16\textfont2=2.7pt \fontdimen17\textfont2=2.7pt
   \mathbf{#1}%
   \resetfontdimens}}
\makeatother

\newcommand{\lsmo}{\chemical{La_{1+x}Sr_{1-x}MnO_4}}

\newcommand{\lsmoe}{\chemical{La_{}Sr_{}MnO_4}}
\newcommand{\lsmofb}{\chemical{La_{\nicefrac{1}{2}}Sr_{\nicefrac{3}{2}}MnO_4}}

\newcommand{\vq}{\chemical{{\bf q}}}

\newcommand{\vQ}{\chemical{{\bf Q}}}

\newcommand{\kommentar}[1]{}

\begin{document}

\title{Spin-wave dispersion in orbitally ordered
La$_{\nicefrac{1}{ 2}}$Sr$_{\nicefrac{3}{ 2}}$MnO$_4$}

\author{D. Senff}
\affiliation{II. Physikalisches Institut, Universit\"at zu K\"oln, Z\"ulpicher Str. 77, D-50937 K\"oln, Germany}

\author{F. Kr\"uger}
\affiliation{Institut f\"ur Theoretische Physik, Universit\"at zu
K\"oln, Z\"ulpicher Str. 77, D-50937 K\"oln, Germany}
\affiliation{Instituut-Lorentz, Universiteit Leiden, P.O. Box
9506, 2300 RA Leiden, The Netherlands}

\author{S. Scheidl}
\affiliation{Institut f\"ur Theoretische Physik, Universit\"at zu K\"oln, Z\"ulpicher Str. 77, D-50937 K\"oln,
Germany}

\author{M. Benomar}
\affiliation{II. Physikalisches Institut, Universit\"at zu K\"oln,
Z\"ulpicher Str. 77, D-50937 K\"oln, Germany}

\author{Y. Sidis}
\affiliation{ Laboratoire L\'eon Brillouin, C.E.A./C.N.R.S., F-91191 Gif-sur-Yvette Cedex, France}

\author{F. Demmel}
\affiliation{Institut Laue-Langevin, Bo\^ite Postale 156, 38042
Grenoble Cedex 9, France}

\author{M. Braden}
\email{braden@ph2.uni-koeln.de}%
\affiliation{II. Physikalisches Institut, Universit\"at zu K\"oln, Z\"ulpicher Str. 77, D-50937 K\"oln, Germany}

\date{\today, \textbf{preprint}}

\pacs{75.20.-m, 71.10.-w, 75.47.Lx, 75.50.Ee, 75.25.+z}

\begin{abstract}

The magnon dispersion in the charge, orbital and spin ordered
phase in \lsmofb \ has been studied by means of inelastic neutron
scattering. We find an excellent agreement with a magnetic
interaction model basing on the CE-type superstructure. The
magnetic excitations are dominated by ferromagnetic exchange
parameters revealing a nearly-one dimensional character at high
energies. The nearest neighbor ferromagnetic interaction in
\lsmofb \ is significantly larger than the one in the metallic
ferromagnetically ordered manganites. The large ferromagnetic
interaction in the charge/orbital ordered phase appears to be
essential for the capability of manganites to switch between
metallic and insulating phases.

\end{abstract}

\maketitle

The phenomenon of colossal magneto-resistivity in the manganites
is only partially explained by the Zener double-exchange
mechanism, the larger part of it appears to arise from the
competition of two states : the metallic ferromagnetically ordered
state on the one side and the insulating one with a cooperative
ordering of charges, orbitals and spins (COS) on the other side
\cite{1,2}. The insulator to metal transition consists in
switching from a phase with long or short-range COS-correlations
into the metallic state where spins are aligned either by external
field or by spontaneous magnetic order. Such interpretation is
strongly supported by studies of the diffuse scattering related
with the COS-correlations: the decrease in electronic resistivity
is found to scale with the suppression of diffuse scattering as
function of either temperature \cite{3,4} or of magnetic field
\cite{5}.

In spite of its eminent relevance for colossal magneto-resistivity
the exact nature of the COS-states in the manganites has not yet
been fully established. The combined COS-ordering has first been
studied in the pioneer work by Wollan and Koehler \cite{6} and by
Goodenough \cite{7} proposing the so-called CE-type arrangement,
which is illustrated in Fig.\ 1a). For half doping, i.e. equal
amounts of the Mn$^{3+}$ and Mn$^{4+}$, there is a checkerboard
arrangement of different charges. In addition the $e_g$-orbitals
at the Mn$^{3+}$ sites form zigzag-chains. The CE-type charge and
orbital arrangement will yield a ferromagnetic interaction in the
zigzag-chains and an antiferromagnetic one in-between. In the
recent literature there is evidence both for \cite{8,9}  and
against \cite{10} this CE-type picture of the COS-state near half
doping. The quantitative structural analysis excludes a full
ordering of charges and orbitals \cite{8,9} which would induce
stronger structural distortions. Recently, a  qualitatively
different scheme was proposed for Pr$_{0.6}$Ca$_{0.4}$MnO$_3$
where charges do not order on the metal sites but on the Mn-O-Mn
bonds forming Zener polarons \cite{10}. Wether this Zener-polaron
picture is applicable for all manganites or whether it is relevant
at all is still under debate \cite{11}.

The magnetic excitations in the ferromagnetic metallic manganites
have been studied by inelastic neutron scattering in many
different compositions \cite{12,13,14,15}, for a recent summary
see reference \cite{15}. At low \vq \ the dispersion is
qualitatively similar in all compounds, it is isotropic and
quadratic, $\omega =Dq^2$, with spin stiffness constants of the
order of   {D$\sim$\unit[150]{meV}}. However at large \vq \ , the
spin-wave dispersion is depending on the exact composition and it
is rather anisotropic. An anomalous softening of the magnons at
the zone-boundaries has been recently attributed to extended
exchange-coupling constants arising from orbital effects
\cite{15}. In view of the large amount of data on the
ferromagnetic phases, it may astonish that there is still no
detailed study on the magnetic excitations in the
antiferromagnetic COS-states. Besides the intrinsic complexity of
the CE-type magnetic ordering, such a study is severely hampered
by the twinning of the manganite crystals in the perovskite
phases. We, therefore, have chosen the layered material \lsmofb \
to study the magnon dispersion in the COS-state. We obtain the
full dispersion of the two magnon branches with lowest energies
which may be satisfactorily described in the CE-type model.

The structural, electronic and magnetic phase diagram of \lsmo \ has been elaborated in references
\cite{16,17,18}. For x=0, \lsmoe , all Mn are three-valent with occupation of   {$(3z^2-r^2)$} $e_g$-orbitals and
spin order antiferromagnetically.
The  half-doped compound, \lsmofb , exhibits the cooperative COS
ordering, which has been studied by various techniques \cite{19,20,21,22,23}. Compared to most perovskite
manganites at half doping, the COS-state in \lsmofb \ appears to be rather stable, only in very high magnetic
fields of the order of \unit[30]{T} the COS-state is suppressed and colossal magnetoresistivity is observed
\cite{24}. For the study of the magnetic excitations in the COS-state this material is nevertheless well suited.

\begin{figure}
\includegraphics*[width=0.45\textwidth]{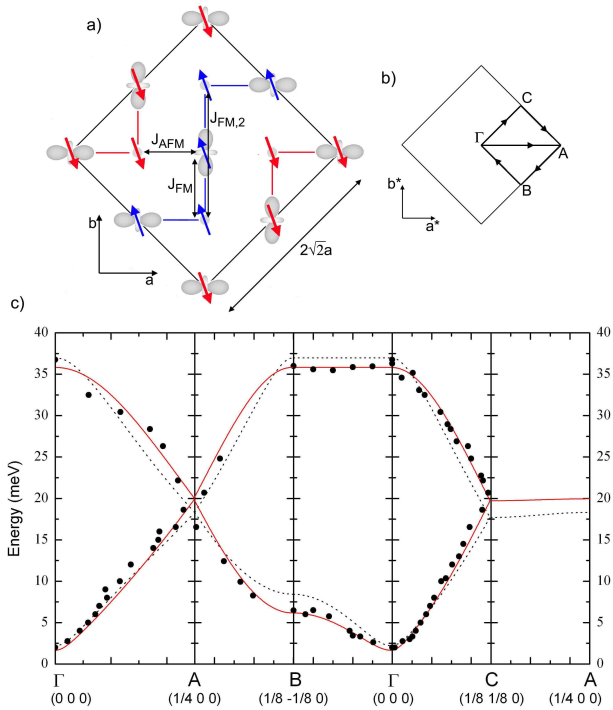}
\caption{(color online) a) Schematic representation of the CE-type
ordering in the ab-plane of half doped
  manganites with the three magnetic interactions parameters described in the text. Notice, that the FM zigzag-chains
  run along the [110]-direction. b) Sketch of the magnetic Brillouin-zone, displaying the high symmetry points $\Gamma=(0,0,0)$,
  $A=(1/4,0,0)$, $B=(1/8,-1/8,0)$, $C=(1/8,1/8,0)$ and the path of the calculated dispersion.
  c) Dispersion of the magnetic excitations in
  \lsmofb \ in a direction parallel [100] ($\Gamma$ to A), perpendicular to the chains ($\Gamma$ to B) and parallel
  to the chains ($\Gamma$ to C). The solid and broken lines give the spin-wave dispersion
  calculated with two parameter sets, see text. }\label{CE-Order}
\end{figure}

The single crystal used in this study was grown by the floating
zone technique as described in \cite{Reutler03} (volume
{$0.6$cm$^3$}, space group $I4/mmm$, $a$=\unit[3.86]{\AA} and
$c$=\unit[12.42]{\AA} at room temperature). Upon cooling we
observe the sequence of structural and magnetic ordering following
the appearance of the respective superstructure reflections in
neutron diffraction experiments. Orbital and charge ordering
within the CE-type picture is related with superstructure
reflections displaced from reciprocal lattice vectors by \vq
=$(\pm\frac{1}{4},\pm\frac{1}{4},0)$ and by
$(\pm\frac{1}{2},\pm\frac{1}{2},0)$ respectively (we use reduced
lattice units of $\frac{2\pi}{a}$ with respect to the $I4/mmm$
cell). The onset of charge/orbital ordering is observed at
  {$T_{CO/OO}=\unit[230]{K}$} in agreement with
references \cite{19,21}. In addition, below $T_{N}=\unit[110]{K}$
  {antiferromagnetic ordering} is evidenced through
magnetic reflections \cite{19}; the magnetic ordering remains,
however, to a large extent two-dimensional in nature in our as
well as in previously studied crystals   {\cite{19}}. For the
determination of the magnon-dispersion the lack of
  {correlations} along the c-direction is irrelevant,
as the magnetic exchange parameters are negligible along this
direction. In the following, we only discuss the layered magnetic
ordering.

Let us illustrate the different propagation vectors with the aid
of the scheme given in Fig.\ 1a). With the orbital ordering, the
nuclear lattice gets orthorhombic with lattice constants of
{$\sqrt{2}a$} along [1,$\bar{1}$,0] and
  {$2\sqrt{2}a$} along [1,1,0]. Note that the
zigzag-chains run along the [1,1,0]-direction. Orbital ordering is
only related to superstructure reflections with \vq
=$\pm(\frac{1}{4},\frac{1}{4},0)$. Considering only the
Mn$^{3+}$-sites the magnetic lattice is orthorhombic too and of
the same size as the structural one but rotated by
  {90\textdegree, $2\sqrt{2}a$} along [1,$\bar{1}$,0]
and   {$\sqrt{2}a$} along [1,1,0]. The Mn$^{3+}$-spins contribute
to magnetic superstructure reflections with \vq
=$\pm(\frac{1}{4},-\frac{1}{4},0)$, for example there is a
contribution at (0.25,0.75,0)=(0.25,-0.25,0) but none at
(0.25,0.25,0). The Mn$^{4+}$-spins do not contribute to neither of
these but to positions with \vq=$\pm(0,0.5,0)$ or
\vq=$\pm(0.5,0,0)$, where the Mn$^{3+}$-spins do not contribute.
The full magnetic cell has to be described in a pseudo-quadratic
lattice with constants   {$2\sqrt{2}a$} along [1,$\bar{1}$,0] and
[1,1,0], as shown in Fig.\ 1a). Due to the twinning in the
orthorhombic COS-phase the arrangement in Fig.\ 1a) (orientation
I) is superposed by the same rotated by
  {90\textdegree} (orientation II) in a sample
crystal. Both twin orientations contribute equally in our sample,
but we will always refer to orientation I for the analysis.

\begin{figure}
  \includegraphics[width=0.4\textwidth]{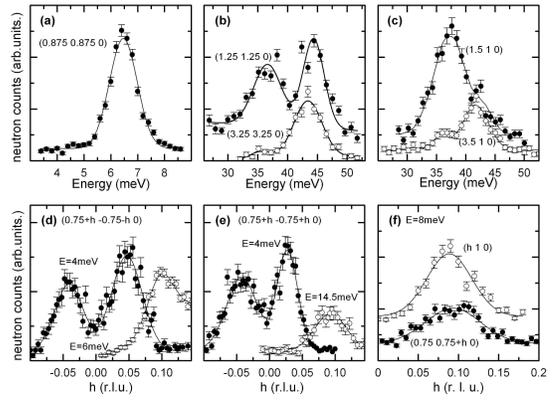}
  \caption{(color online) Raw-data scans to determine the magnon
  dispersion in \lsmofb, symbols denote data and lines fits with gaussians.
a-c) Constant $\bf Q$ scans at
  antiferromagnetic zone-center and zone-boundaries (\ b) and c) were measured with $E_f=\unit[7.37]{meV}$
  and using the copper-monochromator); the different $\bf Q$-dependence
separates magnetic from phononic scattering in b) and c).  d-f) :
Constant
  energy scans for   {different energies} across (0.75,-0.75,0) in
    {$[1,\bar{1},0]$}-direction, i.e. perpendicular to the zigzag-chains d),
and in [1,1,0]-direction e), i.e. parallel to the zigzag-chains. f) Scan aiming at the dispersion along the
[0,1,0]-direction, path {$\bf\Gamma$-{\bf A}}. }
  \label{scans}
\end{figure}

Neutron scattering experiments were performed on the triple-axis
spectrometers $4F$ and  $1T$ at the Orph\'ee reactor in Saclay,
France, and on $IN3$ at the ILL in Grenoble, France. The sample
was mounted with the $[001]$-direction perpendicular to the
scattering plane and cooled to T=15\ K. Monochromatic neutrons
were selected using Bragg-scattering from the $(002)$-reflection
of pyrolytic graphite (PG) or -- at higher incident energies --
the $(111)$-reflection of copper. The final energies of the
scattered neutrons   {were} in most cases fixed to
\unit[14.7]{meV}, to suppress higher order contaminations with the
aid of a PG-filter.

The magnon dispersion in \lsmofb \ was determined by scanning with either \vQ \ or the energy transfer kept
constant. At the antiferromagnetic zone center we find a gap in the magnetic excitation spectrum and a small
energy-splitting of the lowest excitation. The degeneracy of the two transverse magnons appears to be lifted due
to complex magnetic anisotropy terms, see \cite{senff-unp}. Already for
  {q$>$3.9$\cdot10^{-3}$\AA$^{-1}$} away from the zone-centre, no splitting in the magnon
frequencies is resolved anymore. In spite of the twinning of the crystal in the COS-phase we are able to separate
the magnon branches parallel and perpendicular to the zigzag-chains, as only one twin orientation contributes to
a quarter-indexed magnetic superstructure reflection. When going from the antiferromagnetic zone center
(0.75,-0.75,0) along the [1,1,0]-direction one determines the spin-wave dispersion parallel to the zigzag-chains
(see {Fig.\ 2e)}) and when going along the {$[1,\bar{1},0]$}-direction one measures the dispersion perpendicular
to the chains. This behavior is corroborated by the structure factor calculations presented in Fig.\ 3 as
discussed below. The raw-data scans shown in Fig.\ 2 unambiguously   {demonstrate} that the dispersion along the
zigzag-chains is much steeper   {than perpendicular} to them. The magnetic structure has to be considered as a
weak antiferromagnetic coupling of strongly coupled ferromagnetic zigzag-chains. The obtained magnon dispersion
is presented in Fig.\ 1. The branch propagating along the chains, path {$\bf \Gamma$-{\bf C}}, is much steeper
than the branch propagating perpendicular to it, path   {$\bf \Gamma$-{\bf B}}. At the magnetic zone boundaries
{\bf C} and {\bf B} we find magnon energies of   {\unit[19]{meV}} and {\unit[6.5]{meV}}, respectively. At the
point {\bf C} where \vq \ is parallel to the chains, the end-point of the acoustic branch coincides with that of
the lowest optic branch, whereas there is a large gap between these branches along the path {$\bf \Gamma$-{\bf
B}}. The magnon branch along the [100]-direction, path  {$\bf\Gamma$-{\bf A}} at {45\textdegree} to the chains,
exhibits an intermediate dispersion. Finally, all zone-boundary modes connect when passing along the
zone-boundary paths {\bf A}--{\bf B} and {\bf A}--{\bf C}.

We have performed measurements around the quarter-integer indexed
magnetic zone centers as well as around half-integer indexed ones.
As explained above, in elastic scans at these \vQ -values one
strictly measures the scattering contribution of the Mn$^{3+}$-
and that of the Mn$^{4+}$-sites, respectively. This separation
should hold for inelastic scattering at rather low energies as
well. Around these \vQ-values we find exactly the same dispersion,
as it is expected for collective magnons. At finite energies there
is also a significant structure-factor around the integer-indexed
\vQ-values, like (1,0,0); again the dispersion of the modes fully
agrees with the other zones. The   {dispersion shown} in Fig.\ 1
was obtained finally by combining many scans in different magnetic
zones. At energies significantly above the saturation of the
acoustic magnon branch perpendicular to the zigzag-chains, i.e.
  {\unit[6.5]{meV}}, the magnetic interaction
perpendicular to the chains does not play any role
  {anymore} and the magnon dispersion exhibits an
one-dimensional character.


\begin{figure}
\includegraphics[width=0.44\textwidth]{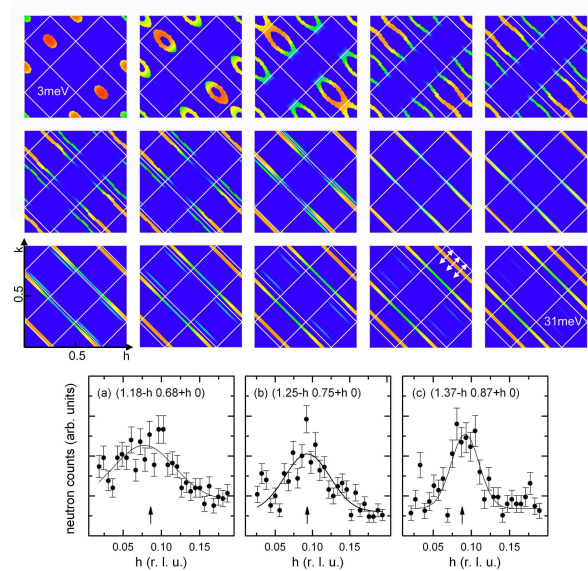}
\caption{(color online) Upper panel: Constant energy cuts through
the calculated spin-wave structure factor $S({\bf Q},\omega)$ with
constant energy-resolution of \unit[2]{meV} within each plot and
with energy-steps of \unit[2]{meV} between adjacent plots, showing
the dispersion and scattering distribution of the lowest magnon
bands. Lower panel : Constant energy scans along the direction
indicated in the   {\unit[29]{meV}} cut above to experimentally
verify the one-dimensional character of the high energy magnetic
scattering. The arrows indicate the expected positions of the
magnon.} \label{Model}
\end{figure}

The spin-wave dispersion has been calculated using the
Holstein-Primakoff transformation with a simple spin-only
hamiltonian illustrated in Fig.\ 1a) (in the sums each pair
appears only once) :


\begin{equation*}
  \begin{split}
    \mathcal{H} = & -\sum\nolimits_{(Mn^{3+},Mn^{4+})_{\vert \vert}}J_{FM}{\bf S_{i}\cdot S_{j}}\\
                  & + \sum\nolimits_{(Mn^{3+},Mn^{4+})_{\perp}}J_{AFM}{\bf S_{i}\cdot S_{j}} \\
                  & - \sum\nolimits_{(Mn^{3+},Mn^{3+})_{nnn\vert \vert}}J_{FM,2}{\bf S_{i}\cdot S_{j}} - \sum{}_{Mn}\Lambda S_z^2.
  \end{split}
\end{equation*}

Details of the calculation can be found in the related work on the
spin excitations in the stripe phases \cite{krueger-scheidl}. The
Mn$^{3+}$ and Mn$^{4+}$-spins were fixed to the values of S=2 and
S=1.5, respectively. Taking into account only the two
nearest-neighbor interactions for Mn$^{3+}$-Mn$^{4+}$-spin
interactions for pairs within and in-between the zigzag-chains,
$J_{FM}$ and $J_{AFM}$, one obtains a good description of the
measured dispersion denoted in the   {Fig.\ 1c)} by broken lines.
However, there remain significant discrepancies; it is impossible
to simultaneously describe the large initial slope of the
spin-wave dispersion along the chains and the relatively lower
zone-boundary frequencies. This behavior implies the relevance of
an additional longer-distance interaction parameter acting along
the ferromagnetic chains. Indeed, a fully satisfactory description
is obtained by including a ferromagnetic interaction for
Mn$^{4+}$-Mn$^{4+}$-spin pairs connected through a Mn$^{3+}$
within a zigzag-chain \cite{note}, see full lines in Fig.\ 1b). We
determine the parameters : $J_{FM}$=  {\unit[9.98]{meV},
$J_{AFM}$=\unit[1.83]{meV}, $J_{FM,2}$=\unit[3.69]{meV}} and an
anisotropy term of  $\Lambda$= {\unit[0.05]{meV}}. This model
predicts the existence of rather flat optical branches around
{\unit[75]{meV}} which, however, could not be observed so far due
to the high phonon signal at these energies. In contrast with the
excellent magnon-dispersion modeling presented above, we do not
find a straightforward description of the observed dispersion
within the Zener-polaron model. The stacking of magnetic dimers
can explain the observed magnetic Bragg-peaks \cite{efremov}, but
one would expect the anisotropy of the dispersion around for
example (0.75,-0.75,0) to be opposite to the experimental finding.
This failure and the excellent spin-wave description obtained
within the CE-type model give strong support for the latter in
\lsmofb .

Fig.\ 3 presents the calculated magnon scattering intensities in the form of constant energy cuts. One can see
how the anisotropic spin-wave cones develop around the magnetic Bragg-peaks with finite structure factor. At
intermediate energies also those magnetic Brillouin-zones contribute where there is no elastic scattering. The
Fig.\ 3 further illustrates that well above the maximum of the acoustic magnon perpendicular to the
zigzag-chains, the system looks like a magnetically one-dimensional system. This character was verified by
special constant-energy scans, see Fig.\ 3. The dominant ferromagnetic coupling is furthermore seen in
experiments upon heating across the charge and orbital ordering transition. The diffuse magnetic scattering as
well as the magnetic fluctuations turn ferromagnetic in character at high temperatures \cite{senff-unp}.

The ferromagnetic interaction in the COS-phase of \lsmofb \ which dominates the spin-wave dispersion is
remarkably large. For example it is about a factor of five larger than the ferromagnetic coupling in LaMnO$_3$
\cite{moussa} acting on two Mn$^{3+}$-sites with an antiferroorbital coupling. $J_{FM}$ is also significantly
larger than the ferromagnetic interaction in the metallic ferromagnetic phases with the highest Curie
temperatures \cite{12,13}. In the Zener double-exchange model a delocalized $e_g$-electron mediates the
ferromagnetic interaction with the neighbors along three orthogonal directions, whereas the orbitally ordered
$e_g$-electron in the CE-type phase focuses the magnetic interaction along one direction. The relevance of the
additional ferromagnetic coupling reminds the same observation in the metallic phases \cite{15} and implies a
similar orbital origin. The dominance of the ferromagnetic interaction in the CE-type phase, which is mediated
through the $e_g$-orbital, appears to be essential for the capability of manganites to switch between the
metallic ferromagnetic and the COS-phases.

In conclusion we have determined the low-energy spin-wave
dispersion in the ordered phase of \lsmofb , which can be
excellently described basing on the CE-type structural model of
charge, orbital and spin ordering.

Work at Universit\"at zu K\"oln was supported by the Deutsche
Forschungsgemeinschaft through the Sonderforschungsbereich 608. We
thank L.P. Regnault for stimulating discussions.

\begin{thebibliography}{28}
\bibitem{note} The magnon dispersion can be described nearly as
well when considering a ferromagnetic coupling between two
Mn$^{3+}$-sites along the zigzag-chain.
\end{thebibliography}

\end{document}